\documentclass[conference]{IEEEtran}
\usepackage{amsmath,amssymb,amsfonts}
\usepackage{graphicx}
\usepackage{textcomp}
\usepackage{xcolor}
\usepackage{booktabs}
\usepackage{cite}
\usepackage{algorithm}
\usepackage{algpseudocode}
\usepackage{comment}
\usepackage{subcaption}
\usepackage{bm}
\newcommand{\diag}{\operatorname{diag}}

\begin{document}

\title{Off-Grid Sparse Target Sensing in OTFS-Based mmWave MIMO Radar With Bayesian Learning}

\author{
\IEEEauthorblockN{Meesam Jafri\textsuperscript{1}, Suraj Srivastava\textsuperscript{2}, Aditya K. Jagannatham\textsuperscript{3}, Marina Petrova\textsuperscript{1,4}, and Emil Bj\"ornson\textsuperscript{1}}
\IEEEauthorblockA{\textsuperscript{1}Department of Communication Systems, KTH, Sweden, \textsuperscript{2} Department of Electronics Engineering, IIT Jodhpur, India \\
\textsuperscript{3} Department of Electrical Engineering, IIT Kanpur, India, \textsuperscript{4} Faculty of Electrical Engineering and Information Technology,\\ RWTH, Germany.\\
Email: meesam@kth.se, surajsri@iitj.ac.in, adityaj@iitk.ac.in, petrovam@kth.se, petrova@mcc.rwth-aachen.de, emilbjo@kth.se}
}
\maketitle
\begin{abstract}
This paper proposes an off-grid sparse Bayesian learning framework for joint delay, Doppler, and angle (DDA) estimation in orthogonal time-frequency-space based millimeter-wave multiple-input multiple-output radar system. Initially, a sparse target DDA parameter estimation model is formulated, which incorporates fractional delay and Doppler shift. Next, an expectation–maximization-based procedure is developed to estimate the radar cross-section (RCS) coefficients and noise variance while refining the DDA grid points.
Subsequently, a joint damped Gauss-Newton update is obtained from a first-order dictionary expansion, retaining posterior RCS uncertainty and coupling among the DDA parameters. The grid point updates are accepted using Armijo backtracking with the exact continuous dictionary. The resulting framework accounts for off-grid DDA values using a coarse initial grid without constructing a densely sampled three-dimensional dictionary. Finally, our simulation results demonstrate the effectiveness of the proposed framework and significantly improved performance over the benchmark schemes.%
\end{abstract}
\begin{IEEEkeywords}
OTFS, radar, mmWave, sparse estimation.
\end{IEEEkeywords}
\vspace{-5pt}
\section{Introduction}
\label{sec:introduction}
Radar sensing is expected to be a core service in 6G wireless communication systems. Deploying radar and communication systems as separate entities could lead to hardware complexity. To ease this, integrated sensing and communication (ISAC) emerged as a key technology for realizing joint radar and communication, which enables a dual-functional radar and communication (DFRC) base station (BS) to support data transmission and radar sensing via shared hardware, spectrum, and waveforms \cite{liu2022integrated}.
Exploiting the mmWave band in DFRC systems is challenging due to the high propagation loss and susceptibility to blockage. At the same time, the small wavelength of mmWave frequencies makes it possible to pack massive antenna arrays into a small area. This facilitates the use of highly focused beamforming to serve multiple users and sense multiple targets simultaneously \cite{zhang2021overview}.

At mmWave frequencies and in high-speed environments, significant Doppler shifts arise, which result in inter-carrier interference and performance degradation in OFDM systems. This motivates the consideration of orthogonal time-frequency-space (OTFS) modulation for high-mobility ISAC systems \cite{hadani2017orthogonal}. OTFS multiplexes information symbols in the delay-Doppler (DD) domain and maps them to the time-frequency (TF) domain for transmission. The DD representation explicitly characterizes the delays and Doppler shifts introduced by the propagation paths. These quantities also represent target range and velocity, which makes OTFS an attractive waveform for ISAC \cite{gaudio2020effectiveness, jafri2024sparse}.

Toward this end, Raviteja \emph{et al.} \cite{raviteja2019orthogonal} proposed matched-filter processing for range and velocity estimation in OTFS-based radar systems. The authors of \cite{jafri2024bayesian} developed Bayesian learning (BL)-based sparse target parameter estimation for OTFS-aided mmWave radar systems, accounting for integer delay and Doppler. The work in \cite{gao2025joint} considers joint delay-Doppler-angular (DDA) parameter estimation in OTFS radar but is limited to an on-grid target scenario. As a further development, authors in \cite{zacharia2023fractional} proposed orthogonal matching pursuit (OMP) with fractional refinement, which estimates integer DD indices followed by a local fractional-dictionary search.
The authors of \cite{liao2023joint} considered the fractional Doppler scenarios while retaining the integer delay for parameter estimation in OTFS systems. Although these approaches account for fractional DD values, estimation over a finite dictionary can still introduce basis mismatch when the true target parameters lie between the DDA grid points. 
To address off-grid parameters, Wei \emph{et al.} \cite{wei2022off} developed sparse Bayesian learning (SBL)-based OTFS channel estimation, wherein DD offsets are learned through expectation-maximization (EM). The  work in \cite{ zhang2024sparse} proposed an off-grid SBL approach that refines fractional Doppler through gradient descent. These channel estimation procedures do not consider joint angular estimation. Furthermore, Dehkordi \emph{et al.} \cite{dehkordi2023beam} developed likelihood-based joint target detection and DDA estimation for OTFS radar with a reduced number of radio frequency chains (RFCs). 

Motivated by these approaches, this paper proposes an off-grid SBL framework for joint RCS coefficient estimation and continuous DDA refinement in OTFS-aided mmWave MIMO radar. The main contributions are summarized as follows:
\begin{itemize}
    \item Initially, a sparse DDA-domain target parameter estimation model is formulated, in which the sensing dictionary incorporates fractional delay and Doppler shift. An EM-based procedure is developed to estimate the RCS coefficients and noise variance while refining the DDA bins.
    \item Next, a first-order dictionary expansion is employed to derive a joint damped Gauss–Newton (GN) update. These updates account for posterior RCS covariance and coupling among the DDA parameters, thus reducing basis mismatch without requiring a finer initial DDA grid.
    \item Subsequently, Armijo backtracking is employed to evaluate the DDA updates using the exact continuous dictionary. Only updates that decrease the posterior-averaged reconstruction error are accepted, thus maintaining the generalized EM property.
    \item Finally, simulation results demonstrate lower RCS coefficient normalized mean squared error (NMSE) and DDA root mean squared error (RMSE) than benchmark schemes. Furthermore, the proposed framework reduces sensitivity to increases in DDA grid intervals.
\end{itemize}
\section{System Model}
\label{SysModel}
Consider a monostatic mmWave MIMO-OTFS radar with $N_{T}$ transmit antennas, $N_{R}$ receive antennas, and
$N_{\mathrm{RF}} \ll \min(N_{T},N_{R})$ RFCs at both transmitter and receiver. The radar transmits an OTFS waveform and detects $L$ targets in the sensing environment. Let $\mathbf{X}_{\mathrm{DD}} \in \mathbb{C}^{M \times N}$ be the DD-domain symbol matrix, where $M$ and $N$ indicate the number of symbols along the delay and Doppler axes, respectively. Also, $\Delta f$ (Hz) and $T$ (seconds) denote the subcarrier spacing and symbol duration, respectively, such that $T \Delta f = 1$. Therefore, the frame duration and bandwidth of the resultant OTFS system are $NT$ and $M\Delta f$, respectively. The transmitter maps the DD domain symbol matrix (DD-SM) $\mathbf{X}_{\text{DD}}$ to the time-frequency (TF) domain symbol matrix (TF-SM) $\mathbf{X}_{\text{TF}} \in \mathbb{C}^{M \times N}$ via the inverse symplectic finite Fourier transform (ISFFT), which can be expressed as $\mathbf{X}_{\text{TF}} = {\mathbf{F}}_M \mathbf{X}_{\text{DD}}  {\mathbf{F}}_N^H$, where $\mathbf{F}_M$ and $\mathbf{F}_N$ denote the discrete Fourier transform (DFT) matrices of order $M$ and $N$, respectively. Thus, the transmit signal matrix $\mathbf{S}' \in \mathbb{C}^{M \times N}$ is given by $\mathbf{S}' = \mathbf{P}_{\mathrm{tx}} \mathbf{X}_{\mathrm{DD}} \mathbf{F}_{N}^{H}$, where  $\mathbf{P}_{\mathrm{tx}} = \mathrm{diag}\{p_{\mathrm{tx}}(pT/M)\}$ collects the pulse-shaping samples~\cite{hadani2017orthogonal}.
\subsection{Radar Channel}
The DDA-domain mmWave MIMO radar channel matrix can be expressed as \cite{jafri2024sparse, raviteja2019orthogonal}
\begin{align}
\mathbf{H}(\tau,\nu, \theta) = \sum_{l=1}^{L} \bar{\mathbf{H}}_{l} \delta(\tau-\tau_{l}) \delta(\nu-\nu_{l}) \delta(\theta-\theta_{l}),
\label{eq:chRadar}
\end{align}
where $\bar{\mathbf{H}}_{l} = \sqrt{N_{T}N_{R}/L}\;\bar{h}_{l}\,
\mathbf{a}_{R}(\theta_{l})\,\mathbf{a}_{T}^{H}(\theta_{l}) \in \mathbb{C}^{N_{R}\times N_{T}}$ and $\tau_l, \nu_{l}, \theta_{l}$ denote the delay, Doppler, and angle, respectively, of the $l$th target. Furthermore, $\bar{h}_{l}$, $\mathbf{a}_{T}(\theta)$ and $\mathbf{a}_{R}(\theta)$ denote the RCS coefficient, transmit and receive array response vectors (ARV), respectively. Since the transmitter and receiver are colocated, the angle of departure equals the angle of arrival. The transmit ARV is given as $\mathbf{a}_{T}(\theta) = \frac{1}{\sqrt{N_{T}}}[1, e^{-j\pi\sin\theta}, \dots, e^{-j\pi(N_{T}-1)\sin\theta}]^{T}$, and similarly for $\mathbf{a}_{R}(\theta)$.
\subsection{Sparse Target Parameter Estimation}
The DFRC BS transmits the $P$ baseband pilot vectors $\mathbf{s}_{z} \in \mathbb{C}^{N_{\mathrm{RF}} \times 1}, 0 \leq z \leq P-1$, and $ \bar{\mathbf{s}}_{z} = \mathbf{F}_{\mathrm{RF}} \mathbf{s}_{z} \in \mathbb{C}^{N_T \times 1}$ is the transmit vector after RF precoding. Concatenate the pilot samples $\mathbf{s}_{z}, 0 \leq z \leq P-1$, to obtain the pilot vector $\mathbf{s}_{a} = \left[ \mathbf{s}_{0}(a),\  \mathbf{s}_{1}(a), \cdots, \mathbf{s}_{P-1}(a)\right]^T \in \mathbb{C}^{P \times 1}$ associated with the $a$th transmit RFC. Let $h_{l} = \sqrt{N_{T}N_{R}/L}\;\bar{h}_{l}$, and define the effective spatial response as $\mathbf{B}(\theta_l) = \mathbf{W}_{\mathrm{RF}}^{H}\mathbf{a}_R(\theta_l)\mathbf{a}_T^{H}(\theta_l)\mathbf{F}_{\mathrm{RF}} \in \mathbb{C}^{N_{\mathrm{RF}} \times N_{\mathrm{RF}}}$, where $\mathbf{W}_\mathrm{RF}$ represents the RF combiner. The matrices $\mathbf{F}_\mathrm{RF}$ and $\mathbf{W}_\mathrm{RF}$ are assumed to be known and remain fixed over the pilot block. After CP removal and RF combining, the pilot output vector $\mathbf{y}_{b} \in \mathbb{C}^{P \times 1}$ across $P$ pilots received at the $b$th receive RFC can be expressed as
\begin{align} 
\mathbf{y}_{b} = \sum_{l=1}^{L} h_{l} \left[\sum_{a=1}^{N_{\text{RF}}} \mathbf{B}_{b,a}(\theta_l) \mathbf{\Pi}\left( \tau_l\right) \mathbf{\Delta}( \nu_l)\mathbf{s}_{a} \right] + \mathbf{n}_{b}, \label{eq:r_l_mat_HB} 
\end{align}
where $\mathbf{B}_{b,a}(\theta_l) = \bigl[\mathbf{W}_{\mathrm{RF}}^{H}\mathbf{a}_R(\theta_l)\mathbf{a}_T^{H}(\theta_l)\mathbf{F}_{\mathrm{RF}}\bigr]_{b,a}$
couples the $a$th transmit RFC to the $b$th receive RFC through the target at angle $\theta_l$ and $\mathbf n_b\in\mathbb C^{P\times1} \sim\mathcal{CN}(\mathbf{0},\sigma^2\mathbf{I}_P)$ is the noise vector at the $b$th receive RFC.
The quantities
$\boldsymbol{\Pi}(\tau_l) \in \mathbb{C}^{P \times P}$ and $\boldsymbol{\Delta}(\nu_l) \in \mathbb{C}^{P \times P}$ account for the fractional DD shift of the $l$th target, respectively. The delay matrix is defined as
\begin{equation*}
\boldsymbol{\Pi}(\tau_l) = \mathbf{F}_P^{H}\,
\boldsymbol{\Omega}^{\kappa_l}\,\mathbf{F}_P, \quad \boldsymbol{\Omega} = \mathrm{diag}\Bigl\{\,e^{-j 2 \pi n / P}\,\Bigr\}_{n=0}^{P-1}
\in \mathbb{C}^{P \times P},
\end{equation*}
where $\mathbf{F}_P \in \mathbb{C}^{P \times P}$ denotes the unitary $P$-point
DFT matrix and $\kappa_l = \tau_l / \Delta \tau$ is the normalized delay with delay resolution $\Delta \tau = 1/(M \Delta f)$. Similarly, the Doppler matrix is defined as $\boldsymbol{\Delta}(\nu_l) =
\mathrm{diag}\Bigl\{\,e^{j 2 \pi \nu_l n T / M}\,\Bigr\}_{n=0}^{P-1}
\in \mathbb{C}^{P \times P}$, where $\nu_l$ is the Doppler shift. When $\kappa_l$ is an integer, $\boldsymbol{\Pi}(\tau_l)$ coincides with the circulant delay matrix used in on-grid OTFS models \cite{jafri2024sparse}.

Stacking the received pilot vectors $\mathbf{y}_b$ across all
receive RFCs into the matrix
$\mathbf{Y} = [\mathbf{y}_1, \dots, \mathbf{y}_{N_{\mathrm{RF}}}] \in \mathbb{C}^{P \times N_{\mathrm{RF}}}$,
and collecting the vectors $\mathbf{s}_a$ into the matrix
$\mathbf{S} = [\mathbf{s}_1, \dots, \mathbf{s}_{N_{\mathrm{RF}}}] \in \mathbb{C}^{P \times N_{\mathrm{RF}}}$,
the received matrix can be written as
\begin{equation}
 \mathbf Y=\sum_{l=1}^{L} h_{l}
  \mathbf{\Pi}\left( \tau_l\right) \mathbf{\Delta}( \nu_l)\mathbf S
  \mathbf B^T(\theta_l)+\mathbf N,
 \label{eq:sm-observation-matrix}
\end{equation}
where $\mathbf{N} = [\mathbf{n}_1, \mathbf{n}_2, \dots, \mathbf{n}_{N_{\mathrm{RF}}}] \in \mathbb{C}^{P \times N_{\mathrm{RF}}}$
collects the noise vectors across all receive RFCs. Finally, vectorizing the output matrix as
$\mathbf y=\operatorname{vec}(\mathbf Y) \in \mathbb C^{D\times1}$, where $D = PN_{\mathrm{RF}}$  yields
\begin{equation}
 \mathbf y=\sum_{l=1}^{L} h_{l}
 \left[\mathbf B(\theta_l)\otimes
       \mathbf{\Pi}\left( \tau_l\right) \mathbf{\Delta}( \nu_l)\right]\mathbf s+\mathbf n,
 \label{MainSysModel}
\end{equation}
where $\mathbf s=\operatorname{vec}(\mathbf S)$ and $\mathbf n=\operatorname{vec}(\mathbf N) \in \mathbb C^{D\times1}$.
For sparse estimation, we divide the scattering scene into $G_{\tau}$ delay, $G_{\nu}$ Doppler, and $G_{\theta}$ angle bins, giving $G = G_{\tau} G_{\nu} G_{\theta}$ grid points. The sparse 3D DDA-domain mmWave MIMO equivalent representation of the radar channel can be formulated as
\begin{align} \label{sparsity1}
\mathbf {H}(\tilde{\tau},\tilde{\nu},\tilde{\theta}) = & \sum _{i=0}^{G_{\tau }-1} \sum _{j=0}^{G_{\nu}-1} \sum _{k=0}^{G_{\theta}-1} \mathbf{H}_{i,j,k}\delta (\tilde{\tau} -\tau _{i}) \delta (\tilde{\nu} -\nu_{j})  \delta (\tilde{\theta} -\theta_{k}),
\end{align} 
where \(\mathbf{H}_{i,j,k}\) represents the effective RCS matrix for the target present in the \((i,j,k)\) DDA bin.
The sparse parameter estimation model of \eqref{MainSysModel} using \eqref{sparsity1} can be rewritten as
\begin{align} 
\mathbf{y} = \sum _{i=0}^{G_{\tau }-1} \sum _{j=0}^{G_{\nu}-1} \sum _{k=0}^{G_{\theta}-1} {h}_{i,j,k} \boldsymbol{\phi}_{i,j,k}  + \mathbf{n}, \label{eq:r_l_mat_HB_1} 
\end{align} 
where $\boldsymbol{\phi}_{i,j,k} = \left[\mathbf B(\theta_k)\otimes
\mathbf{\Pi}\left( \tau_i\right) \mathbf{\Delta}( \nu_j) \right] \mathbf{s} \in \mathbb{C}^{D \times 1}$ and ${h}_{i,j,k}$ represents the RCS coefficient corresponding to the $i$th delay, $j$th Doppler and $k$th angular bins. Concatenating the quantities $\boldsymbol{\phi}_{i,j,k}, 0 \leq i \leq G_{\tau}-1, 0 \leq j \leq G_{\nu}-1, 0 \leq k \leq G_{\theta}-1$, to obtain the dictionary matrix $\mathbf{\Phi} \in \mathbb{C}^{D \times G}$. Thus, the sparse parameter estimation model can be expressed as
\begin{align}
    \mathbf{y} = \mathbf{\Phi} \mathbf{h} + \mathbf{n}, \label{eq:sparse_model}
\end{align}
where the quantity $\mathbf{h}  \in \mathbb{C}^{G \times 1}$ denotes the stacked coefficients $h_{i,j,k}, \forall\ i,j,k$, in a suitable indexing order.
\section{Off-Grid Sparse Bayesian Learning}
The on-grid model \eqref{eq:sparse_model} assumes that each target's delay, Doppler, and angle coincide with grid points. Because these parameters are continuous, the true values generally lie off the grid. A finer DDA grid reduces basis mismatch but increases the dictionary size and the correlation between neighboring dictionary columns. This section develops an off-grid sparse Bayesian learning framework that treats the DDA coordinates as continuous parameters. 
Let $\boldsymbol{\eta}_{g} = (\tau_{g}, \nu_{g}, \theta_{g})^{T} \in \mathbb{R}^{3}$ denotes the $g$th grid point and $\boldsymbol{\eta} = [\boldsymbol{\eta}_{1}^{T}, \dots, \boldsymbol{\eta}_{G}^{T}]^{T} \in \mathbb{R}^{3G}$. Write $\mathbf\Phi(\boldsymbol\eta)=[\boldsymbol\phi(\boldsymbol\eta_1),\dots,\boldsymbol\phi(\boldsymbol\eta_G)]$. The derivatives of the $g$th column with respect to its delay, Doppler, and angle are
\begin{align}
\boldsymbol{\xi}_{\tau,g}
&= \bigl[\mathbf{B}(\theta_{g}) \otimes
\boldsymbol{\Pi}'(\tau_{g})\boldsymbol{\Delta}(\nu_{g})\bigr]\mathbf{s},
\label{eq:xi_tau} \\[2pt]
\boldsymbol{\xi}_{\nu,g}
&= \bigl[\mathbf{B}(\theta_{g}) \otimes
\boldsymbol{\Pi}(\tau_{g})\boldsymbol{\Delta}'(\nu_{g})\bigr]\mathbf{s},
\label{eq:xi_nu} \\[2pt]
\boldsymbol{\xi}_{\theta,g}
&= \bigl[\mathbf{B}'(\theta_{g}) \otimes
\boldsymbol{\Pi}(\tau_{g})\boldsymbol{\Delta}(\nu_{g})\bigr]\mathbf{s},
\label{eq:xi_theta}
\end{align}
where $\boldsymbol{\Pi}'(\tau) = \frac{\partial \boldsymbol{\Pi}(\tau)}{\partial \tau}$, $\boldsymbol{\Delta}'(\nu) = \frac{\partial \boldsymbol{\Delta}(\nu)}{\partial \nu}$ and $\mathbf{B}'(\theta) = \frac{\partial \mathbf{B}(\theta)}{\partial \theta}$.

The SBL framework initially assigns the parameterized Gausssian prior below to the unknown channel vector $\mathbf{h}$
\begin{align}
p(\mathbf {h}; \mathbf \Gamma)=\prod _{g=1}^{G}(\pi \gamma _g)^{-1} \exp \Bigg (-\displaystyle \frac{|\mathbf h(g)|^2}{\gamma _g}\Bigg),
\end{align}
where the hyperparameter $\gamma _i,\forall \ 1 \leq i \leq G$ denotes the prior variance corresponding to the $i$th component of $\mathbf{h}$ and $ \ \mathbf {\Gamma } = \mathrm{diag}(\gamma _1, \gamma _2, \ldots, \gamma _{G})\in \mathbb {R}^{G \times G}$ is the diagonal matrix of hyperparameters.  From the prior assignment above, it can be observed that the channel component $\mathbf {h}(i)\rightarrow 0$ as the associated hyperparameter $\gamma _i \rightarrow 0$. Thus, estimation of $\mathbf{h}$ reduces to estimating the associated hyperparameter vector $\boldsymbol{\gamma }=[\gamma _1, \gamma _2, \ldots, \gamma _{G}]^T$. The optimization objective for maximum-likelihood (ML) estimation of the hyperparameter vector $\boldsymbol{\gamma}$ is determined as
\begin{align}
\log \ p(\mathbf {y};\boldsymbol\Theta) = -D\log\pi - \log \left(\det \left(\boldsymbol{\Sigma }_{\mathbf {y}}\right)\right) - \mathbf {y}^H \boldsymbol{\Sigma }_{\mathbf {y}}^{-1} \mathbf {y}, \label{eq:evidence}
\end{align}
where $\boldsymbol\Theta=(\boldsymbol\gamma,\boldsymbol\eta,\sigma^2)$ and ${\boldsymbol{\Sigma }}_{\mathbf y}= \sigma^2 \mathbf {I}_D +  \mathbf{\Phi} {\mathbf {\Gamma }} \mathbf{\Phi}^H \in \mathbb {C}^{D \times D}$. Direct maximization of~\eqref{eq:evidence} is a non-convex problem. Hence, we employ the EM framework. The hyperparameters and noise variance are constrained as $\gamma_g \geq \gamma_{\min}>0$, $g=1,\ldots,G$, and $\sigma^2 \geq \sigma_{\min}^2>0$, respectively. These lower bounds ensure that the prior covariance matrix $\boldsymbol{\Gamma}$ remains non-singular and prevent the noise variance from collapsing to zero.
\subsection{Expectation step (E-step)}
\label{sec:offgrid_em}
Let $\mathbf{\Phi}^{t} = \mathbf{\Phi}(\boldsymbol{\eta}^{(t)})$ and $\mathbf{\Gamma }_{(t)} = \mathrm{diag}(\boldsymbol{\gamma }^{(t)})$ denote the dictionary and prior covariance at iteration $t$. The E-step evaluates average log-likelihood
\begin{equation}
\mathcal{L}(\boldsymbol\Theta\mid\boldsymbol\Theta^{(t)})
=\mathbb E_{\mathbf h\mid\mathbf y;\boldsymbol\Theta^{(t)}}
\left[\log p(\mathbf y\mid\mathbf h;\boldsymbol\eta,\sigma^2)
+\log p(\mathbf h;\boldsymbol\Gamma)\right],
\label{eq:Qdefinition}
\end{equation}
under the Gaussian posterior $p(\mathbf h\mid\mathbf y;\boldsymbol\Theta^{(t)})
=\mathcal{CN}(\mathbf h;\boldsymbol\mu_t,\boldsymbol\Sigma_t)$,
whose covariance and mean are given by
\begin{align}
\boldsymbol\Sigma_t
&=\left(\boldsymbol\Gamma_t^{-1}
+\sigma_t^{-2}\boldsymbol\Phi_t^H
\boldsymbol\Phi_t\right)^{-1}\in\mathbb{C}^{G\times G},\label{eq:Sigma}\\
\boldsymbol\mu_t
&=\sigma_t^{-2}\boldsymbol\Sigma_t\boldsymbol\Phi_t^H
\mathbf y\in\mathbb{C}^{G\times1}.
\label{eq:mu}
\end{align}
The posterior second moment is given as $\mathbf M_t=\boldsymbol\mu_t\boldsymbol\mu_t^H+\boldsymbol\Sigma_t
\in\mathbb{C}^{G\times G}$, and the posterior averaged reconstruction error is
\begin{equation}
\begin{split}
\mathcal{E}^{(t)}(\boldsymbol\eta)
&=\mathbb E
\|\mathbf y-\boldsymbol\Phi(\boldsymbol\eta)\mathbf h\|^2\\
&=\|\mathbf y-\boldsymbol\Phi(\boldsymbol\eta)\boldsymbol\mu_t\|^2
+\mathrm{Tr}\!\left[\boldsymbol\Phi(\boldsymbol\eta)
\boldsymbol\Sigma_t\boldsymbol\Phi^H(\boldsymbol\eta)\right].
\end{split}
\label{eq:R}
\end{equation}
Substituting $\mathbf M_t$ and~\eqref{eq:R}
into~\eqref{eq:Qdefinition} gives
\begin{equation}
\begin{split}
\mathcal{L}(\boldsymbol\Theta\mid\boldsymbol\Theta^{(t)})
={}&c-D\log\sigma^2-\frac{\mathcal{E}^{(t)}(\boldsymbol\eta)}{\sigma^2}\\
&-\sum_{g=1}^{G}\left(\log\gamma_g+
\frac{[\mathbf M_t]_{g,g}}{\gamma_g}\right),
\end{split}
\label{eq:L}
\end{equation}
where $c=-(D+G)\log\pi$. The moments $\boldsymbol\mu_t$,
$\boldsymbol\Sigma_t$, and $\mathbf M_t$ remain fixed throughout the following M-step.
\subsection{Maximization (M)-step}
The M-step updates $\boldsymbol\gamma$, $\boldsymbol\eta$, and
$\sigma^2$ by increasing the same objective in~\eqref{eq:L}.
The three updates are described below.
\subsubsection{Hyperparameter update}
The terms involving $\boldsymbol\gamma$ separate across the $G$ coefficients. Consequently,
\begin{equation}
\gamma_g^{(t+1)}
=\arg\max_{\gamma_g\geq\gamma_{\min}}
\left[-\log\gamma_g-\frac{[\mathbf M_t]_{g,g}}{\gamma_g}\right].
\label{eq:upsilonopt}
\end{equation}
Setting the derivative to zero yields $-1/\gamma_g+[\mathbf M_t]_{g,g}/\gamma_g^2=0$. Thus, the constrained update is
\begin{equation}
\gamma_g^{(t+1)}
=\max\!\left\{\gamma_{\min},
|\mu_{t,g}|^2+[\boldsymbol\Sigma_t]_{g,g}\right\},
\qquad g=1,\ldots,G,
\label{eq:upsilon}
\end{equation}
where $\mu_{t,g}=[\boldsymbol\mu_t]_g$. This is the single-vector SBL hyperparameter update.
\subsubsection{Off-grid DDA update}
During the M-step, the posterior moments \(\boldsymbol{\mu}_t\) and \(\boldsymbol{\Sigma}_t\), obtained from the E-step, remain fixed. The only term in \eqref{eq:L} that depends on $\boldsymbol{\eta}$ is $-\frac{\mathcal{E}^{(t)}(\boldsymbol{\eta})}{\sigma_t^2}$. Thus, maximizing the EM objective with respect to \(\boldsymbol{\eta}\) is equivalent to solving
\begin{align}
\boldsymbol{\eta}^{(t+1)} = \arg\min_{\boldsymbol{\eta}\in\mathcal B} \mathcal{E}^{(t)}(\boldsymbol{\eta}), \label{etaMinimize}
\end{align}
where $\mathcal{B}$ denotes the feasible set that keeps every grid point inside the DDA bin of its initial grid point. Let $\boldsymbol\eta_g^{(0)}=(\tau_g^{(0)},\nu_g^{(0)},\theta_g^{(0)})^T$ denote the initial coordinates of the \(g\)th grid point and let $h_i$ be the grid spacing of
coordinate $i\in\{\tau,\nu,\theta\}$. Furthermore, $[\eta_{i,\min},\eta_{i,\max}]$ denote the search interval of coordinate $i$ that bounds the region of the DDA space under consideration. It contains the initial grid, and the refined coordinates of the grid points are not allowed to leave it. The interval of coordinate $i$ of grid point $g$ is defined as $\underline{\eta}_{i,g} = \max\left\{ \eta_{i,\min}, \eta_{i,g}^{(0)}-\frac{h_i}{2} \right\}$ and $\overline{\eta}_{i,g} = \min\left\{ \eta_{i,\max},
\eta_{i,g}^{(0)}+\frac{h_i}{2} \right\}$.
Accordingly, the feasible set of grid point $g$ is
\begin{align}
\mathcal B_g = \left\{ \boldsymbol{\eta}_g\in\mathbb R^3: \underline{\eta}_{i,g} \leq \eta_{i,g} \leq \overline{\eta}_{i,g}, \quad i\in\{\tau,\nu,\theta\} \right\},
\end{align}
and the feasible set for all $G$ grid points is $\mathcal B = \mathcal B_1\times\mathcal B_2\times\cdots\times\mathcal B_G \subset\mathbb R^{3G}$. 
Thus, every grid point may move continuously in delay, Doppler, and angle, but remains within the DDA bin associated with its initial grid point. These bounds prevent a grid point from moving across several grid bins during a single refinement.
Since \(\boldsymbol{\Phi}(\boldsymbol{\eta})\) depends nonlinearly on $\boldsymbol{\eta}$, obtaining the exact solution of \eqref{etaMinimize} is intractable. Thus, we perform one damped GN step described in Section III-D. The resulting step is accepted only when it satisfies $\mathcal{E}^{(t)}\!\left(\boldsymbol{\eta}^{(t+1)}\right) \leq \mathcal{E}^{(t)}\!\left(\boldsymbol{\eta}^{(t)}\right)$.
The accepted step increases the expected log-likelihood which makes it a valid generalized EM update, without exact minimization of \eqref{etaMinimize}.
\subsubsection{Noise update}
After updating the DDA coordinates, the noise variance is updated using the same E-step moments
\begin{equation}
\sigma_{t+1}^{2} = \max\Bigl\{\sigma_{\min}^{2},\,
\frac{\mathcal{E}^{(t)}(\boldsymbol{\eta}^{(t+1)})}{D}\Bigr\}.
\label{eq:noise_update}
\end{equation}
\subsection{Joint DDA Refinement}
\label{sec:offgrid_refine}
We now update the grid points $\boldsymbol{\eta}$ inside the EM
loop. Towards this, let $p_{t,g} = [\mathbf{M}_{t}]_{g,g}$ denote the posterior expected power of the $g$th grid point. We select at most $K_{\mathrm{act}}$ grid points with the largest $p_{t,g}$ that also satisfy $p_{t,g} \ge \rho \max_{g'} p_{t,g'}$, where $0 < \rho < 1$. Let $\mathcal{S}_{t}$ denote the index set of the selected grid points and $K_{t} = |\mathcal{S}_{t}|$. Only the $3K_{t}$ DDA parameters of the selected grid points are refined. All $G$ dictionary columns remain in the reconstruction error.

Next, arrange the selected DDA parameters in order by grid-point index, with delay, Doppler, and angle consecutive for each grid points. For $i$th selected DDA parameter, let $\mathbf{d}_{i} \in \mathbb{C}^{D \times 1}$ denote the derivative of the dictionary column with respect to the corresponding DDA parameter, that is, $\boldsymbol{\xi}_{\tau, g_{i}}$, $\boldsymbol{\xi}_{\nu, g_{i}}$, or $\boldsymbol{\xi}_{\theta, g_{i}}$ as defined in \eqref{eq:xi_tau}--\eqref{eq:xi_theta}. Linearizing $\boldsymbol\Phi(\boldsymbol\eta)$ to first order around $\boldsymbol\eta^{(t)}$ with these columns gives the gradient $\mathbf{b}_{t}\in\mathbb R^{3K_t}$ of $\mathcal E_t(\boldsymbol\eta)/\sigma_t^2$ and its GN Hessian $\mathbf H_t\in\mathbb R^{3K_t\times3K_t}$ \cite{nocedal2006numerical} as
\begin{align}
[\mathbf{b}_{t}]_{i} &= \frac{2}{\sigma_{t}^{2}}\,
\Re\Bigl\{-\mu_{t,g_{i}}^{*}\,\mathbf{d}_{i}^{H}\mathbf{r}_{t}
+ \mathbf{d}_{i}^{H}\boldsymbol{\Phi}_{t}
[\boldsymbol{\Sigma}_{t}]_{:, g_{i}}\Bigr\},
\label{eq:gradient}\\
[\mathbf{H}_{t}]_{i,j} &= \frac{2}{\sigma_{t}^{2}}\,
\Re\bigl\{(\mathbf{d}_{i}^{H}\mathbf{d}_{j})
[\mathbf{M}_{t}]_{g_{j}, g_{i}}\bigr\},
\label{eq:hessian}
\end{align}
where $\mathbf{r}_{t} = \mathbf{y} - \boldsymbol{\Phi}_{t}\boldsymbol{\mu}_{t}$ is the current reconstruction error of the posterior mean.
The posterior second moment $\mathbf{M}_{t}$ includes both the posterior mean and the posterior covariance, so the Hessian reflects both the estimated magnitudes of the gains and their uncertainty. The off-diagonal entries capture the coupling between the delay, Doppler, and angle parameters of the same grid point, and between different grid points.
\subsubsection{Gauss-Newton Update} Since delay, Doppler, and angle have different physical units, their updates are normalized by the corresponding grid intervals before imposing damping and step-size constraints. To handle this, we define the scaling matrix as $\mathbf{D}_h=\diag(h_\tau,h_\nu,h_\theta)\in\mathbb{R}^{3\times3}$ and the displacement of active grid points $g$ can be written as $\Delta\boldsymbol\eta = \mathbf D_t\mathbf u$, where $\mathbf{D}_{t} = \mathbf{I}_{K_{t}} \otimes \mathbf{D}_{h}
\in \mathbb{R}^{3K_{t} \times 3K_{t}}$ and $\mathbf{u} \in \mathbb{R}^{3K_{t} \times 1}$ denote a dimensionless increment obtained by solving the strictly convex quadratic program
\begin{equation}
\begin{aligned}
\mathbf{u}_{t} = \arg\min_{\mathbf{u}} \quad
&(\mathbf{D}_{t}^{T}\mathbf{b}_{t})^{T}\mathbf{u}
+ \tfrac{1}{2}\mathbf{u}^{T}
(\mathbf{D}_{t}^{T}\mathbf{H}_{t}\mathbf{D}_{t} + \lambda_{t}\mathbf{I})
\mathbf{u} \\
\text{s.t.} \quad
&\underline{\boldsymbol{\eta}}_{\mathcal{S}_{t}}
\le \boldsymbol{\eta}_{\mathcal{S}_{t}}^{(t)} + \mathbf{D}_{t}\mathbf{u}
\le \overline{\boldsymbol{\eta}}_{\mathcal{S}_{t}}, \\
&\|\mathbf{u}\|_{\infty} \le u_{\max},
\end{aligned}
\label{eq:QP}
\end{equation}
where $\lambda_t>0$ is the damping parameter \cite{dennis1996numerical} and where $u_{\max} > 0$ limits the displacement of each DDA parameter per iteration.
If the unconstrained solution satisfies the prescribed DDA parameter bounds and the step-size constraint, the update is given by $\mathbf{u}_{t} = -(\mathbf{D}_{t}^{T}\mathbf{H}_{t}\mathbf{D}_{t}
+ \lambda_{t}\mathbf{I})^{-1}\mathbf{D}_{t}^{T}\mathbf{b}_{t}$. Otherwise, $\mathbf{u}_{t}$ is obtained by solving the quadratic problem in \eqref{eq:QP} subject to these constraints.
\subsubsection{Armijo Backtracking for DDA Grid Refinement}
\label{sssec:line_search}
Define the full grid-point refinement direction as
\begin{equation}
\Delta\boldsymbol{\eta}_{g}^{(t)} =
\begin{cases}
\mathbf{D}_h\mathbf{u}_g^{(t)}, & g \in \mathcal{S}_t, \\[2pt]
\mathbf{0}, & g \notin \mathcal{S}_t,
\end{cases}
\label{eq:full_direction}
\end{equation}
so that only the active grid points receive a nonzero displacement.
Starting with $\alpha = 1$, test
$\alpha \in \{1, \beta, \beta^{2}, \dots\}$, where $0 < \beta < 1$, and accept the first value satisfying the Armijo condition \cite{nocedal2006numerical, armijo1966minimization}
\begin{equation}
\begin{aligned}
\frac{1}{\sigma_t^2}\mathcal{E}_{t}\bigl(\boldsymbol{\eta}^{(t)} + \alpha \Delta\boldsymbol{\eta}^{(t)}\bigr)
\le \frac{1}{\sigma_t^2}\mathcal{E}_{t}(\boldsymbol{\eta}^{(t)})  + c_S \alpha \sum_{g \in \mathcal{S}_t}
(\mathbf{b}_g^{(t)})^{T}\Delta\boldsymbol{\eta}_g^{(t)},
\end{aligned}
\label{eq:armijo}
\end{equation}
where $0 < c_S < 1$, and $\mathbf b_g^{(t)}$ collects the entries of $\mathbf b_t$ of grid point $g$. Each trial is evaluated with the exact continuous dictionary $\boldsymbol{\Phi}(\boldsymbol{\eta})$. The accepted grid point update is
\begin{equation}
\boldsymbol{\eta}_{g}^{(t+1)} = \boldsymbol{\eta}_{g}^{(t)}
+ \alpha_t \Delta\boldsymbol{\eta}_{g}^{(t)},
\qquad g = 1, \dots, G.
\label{eq:update}
\end{equation}
If the quadratic subproblem does not yield a descent direction, or if no trial value satisfies \eqref{eq:armijo} within the backtracking budget, we set $\alpha_t = 0$ and retain the current DDA grid. The hyperparameter and noise variance updates still proceed.

At convergence $t=t_c$, the estimate $\hat{\mathbf{h}}$ of the RCS coefficient is obtained from the posterior mean $\boldsymbol{\mu}_{t_c}$, and the DDA parameter estimates are obtained from the refined grid positions $\boldsymbol{\eta}^{(t_c)}$.  Since only $L$ DDA bins are non-zero, we identify the target grid points using the posterior expected power $p_{T,g}$.  Let $g_{1}, \dots, g_{L}$ denote the indices of the $L$ grid points with largest $p_{T,g}$, the estimated DDA parameters and RCS coefficients of $L$ targets are given as 
\begin{equation}
\hat{\tau}_{l} = \tau_{g_{l}}^{(t_c)}, \quad
\hat{\nu}_{l} = \nu_{g_{l}}^{(t_c)}, \quad
\hat{\theta}_{l} = \theta_{g_{l}}^{(t_c)}, \quad
\hat{h}_{l} = \mu_{t_c, g_{l}}.
\label{eq:estimates}
\end{equation}
The range and radial velocity follow from $\hat{r}_{l} = c\,\hat{\tau}_{l}/2$ and $\hat{v}_{l} = \lambda_{c}\,\hat{\nu}_{l}/2$, where $c$ is the propagation speed and $\lambda_{c}$ is the carrier wavelength.
The complete procedure is summarized in Algorithm
\ref{alg:offgrid}.
\begin{algorithm}[t]
\caption{Off-grid SBL-based DDA estimation}
\label{alg:offgrid}
\begin{algorithmic}[1]
\State \textbf{Input:} Observation vector $\mathbf{y}$, maximum EM iterations $S_{\mathrm{max}}$, stopping parameter $\epsilon_s$
\State \textbf{Initialize:} $\boldsymbol{\Phi}_{0} = \boldsymbol{\Phi}(\boldsymbol{\eta}^{(0)})$;
$\boldsymbol{\gamma}^{(0)} = \frac{\|\mathbf{y}\|_{2}^{2}}{\|\boldsymbol{\Phi}_{0}\|_{F}^{2}}, g = 1, \dots, G,$ and $\sigma_{0}^{2}= \frac{\|\mathbf{y}\|_{2}^{2}}{D}$; $t = 0$
\While{stopping criterion \eqref{eq:stop} is not met and
$t < S_{\max}$}
    \State Form the continuous dictionary $\mathbf{\Phi}_{(t)} =
    \mathbf{\Phi}(\boldsymbol{\eta}^{(t)})$
    \State \textbf{E-step:} compute $\boldsymbol{\Sigma}_{(t)}$,
    $\boldsymbol{\mu}_{(t)}$, and $\mathbf{M}_{(t)}$ 
    \State \textbf{M-step:} i) update $\gamma_{g}$ via
    \eqref{eq:upsilon}
    \State \textbf{Active set:} select $\mathcal{S}_{(t)}$ from the posterior
    powers $p_{t,g}$
    \State \textbf{DDA parameter update:}
    \State \quad Evaluate derivative columns
    \eqref{eq:xi_tau}--\eqref{eq:xi_theta}
    \State \quad Form gradient $\mathbf{b}_{t}$ and Hessian
    $\mathbf{H}_{t}$ via \eqref{eq:gradient}, \eqref{eq:hessian}
    \State \quad Solve \eqref{eq:QP}
    for $\{\mathbf{u}_{g}^{(t)}\}_{g \in \mathcal{S}_{t}}$
    \State \quad Update $\boldsymbol{\eta}^{(t+1)}$
    via \eqref{eq:update}
    \State \textbf{Noise update:} update $\sigma_{t+1}^{2}$ via
    \eqref{eq:noise_update}
    \State $t \leftarrow t + 1$
\EndWhile
\State \textbf{Output:} DDA parameters and RCS coefficients.
\end{algorithmic}
\end{algorithm}
\begin{figure*}[!t]
    \centering
    \begin{subfigure}[t]{0.24\textwidth}
        \centering
        \includegraphics[width=\linewidth]{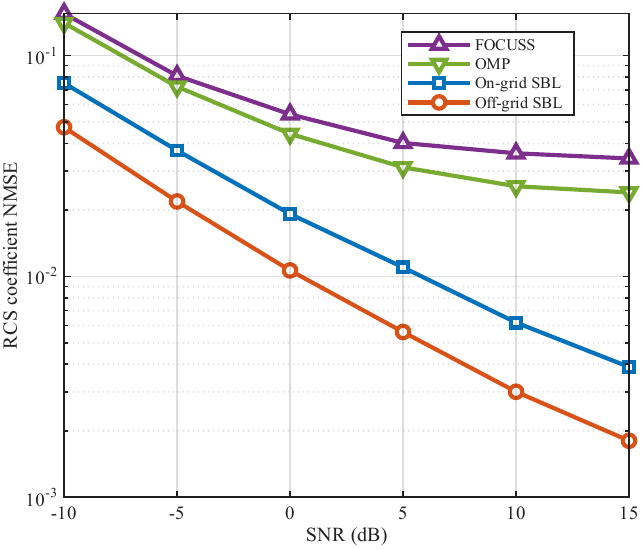}
        \caption{RCS NMSE.}
        \label{fig:rcs}
    \end{subfigure}
    \hfill
    \begin{subfigure}[t]{0.24\textwidth}
        \centering
        \includegraphics[width=\linewidth]{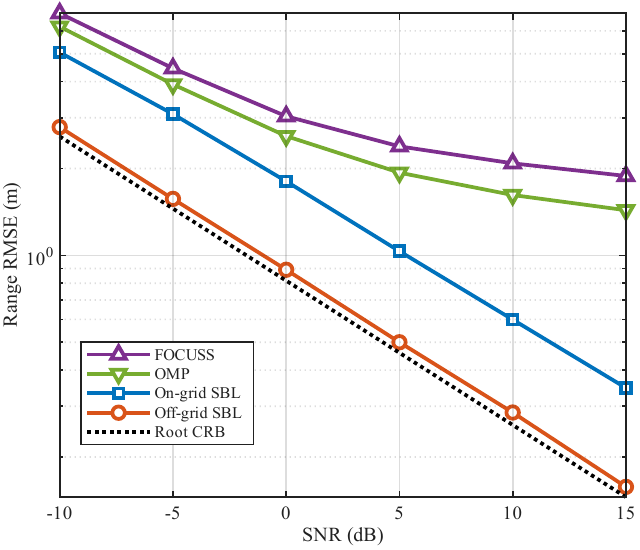}
        \caption{Range RMSE.}
        \label{RangeRMSE}
    \end{subfigure}
    \hfill
    \begin{subfigure}[t]{0.24\textwidth}
        \centering
        \includegraphics[width=\linewidth]{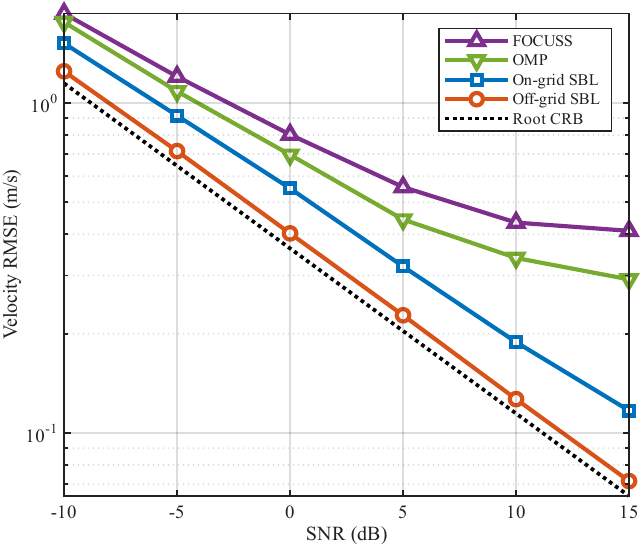}
        \caption{Velocity RMSE}
        \label{VelocityRMSE}
    \end{subfigure}
    \hfill
    \begin{subfigure}[t]{0.24\textwidth}
        \centering
        \includegraphics[width=\linewidth]{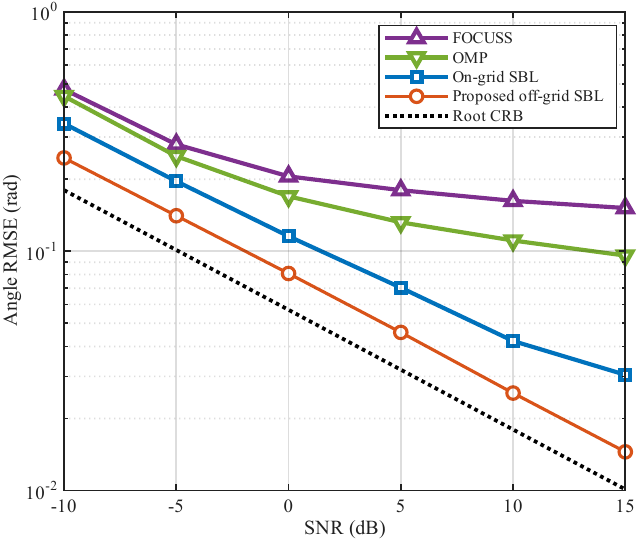}
        \caption{Angle RMSE}
        \label{AngleRMSE}
    \end{subfigure}
    \caption{NMSE and RMSE versus SNR for estimation of RCS coefficient, range, velocity and angle.}
    \label{fig:rmse}
    \vspace{-4mm}
\end{figure*}
The stopping criterion is met when both
\begin{equation}
\frac{|\mathcal{L}_{(t+1)} - \mathcal{L}_{(t)}|}{1 + |\mathcal{L}_{(t)}|} \le \epsilon_{l}
\qquad\text{and}\qquad
\|\alpha_{(t)}\mathbf{u}_{(t)}\|_{\infty} \le \epsilon_{\eta},
\label{eq:stop}
\end{equation}
where $\mathcal{L}_{(t)}$ is the evidence \eqref{eq:evidence} at iteration $t$. 
\begin{table}[t]
\caption{Simulation parameters.}\label{tab:setup}
\centering\footnotesize
\begin{tabular}{@{}ll@{}}\toprule
Parameter & Value\\\midrule
Carrier frequency / subcarrier spacing & $30$ GHz / $120$ kHz\\
OTFS dimensions & $M=N=32$\\
$N_T$ / $N_R$ / RFCs & $32$ / $32$ / $3$\\
Number of pilots / targets & $P=256$ / $L=3$\\
Initial normalized delay grid & $0:0.5:6$\\
Initial normalized Doppler grid & $-3:0.5:3$\\
Initial angle grid & $-7.2^\circ:0.6^\circ:7.2^\circ$\\
Maximum EM iterations ($S_{\mathrm{max}}$) & $150$\\\bottomrule
\end{tabular}
\end{table}
\section{Simulation Results}
\label{sec:sims}
\begin{figure}
    \centering
    \includegraphics[width=0.5\linewidth]{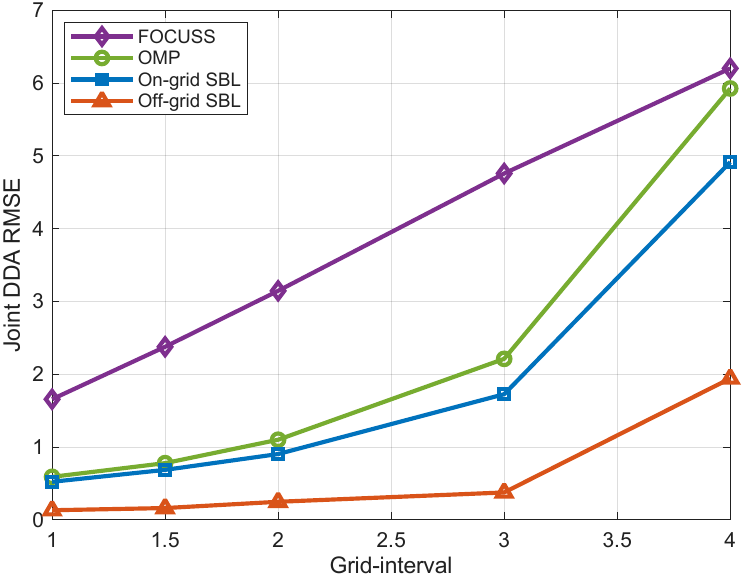}
   \caption{Joint DDA RMSE versus dictionary grid interval $q$ at SNR $=10$dB.}
\label{fig:grid}
\vspace{-5mm}
\end{figure}
For characterizing the performance of Off-grid SBL parameter estimation in a mmWave MIMO OTFS system, simulation parameters are listed in Table~\ref{tab:setup}. The $P = 256$ pilot symbols are randomly generated from the QPSK alphabet. The RF precoder and combiner are selected from the DFT codebook. For each Monte Carlo realization, \(L=3\) point targets are generated within the sensing region. Their delay, Doppler, and angle parameters are obtained by randomly selecting interior points of the initial DDA grid and adding independent uniformly distributed offsets within
 $[-0.45h_{i,0},0.45h_{i,0}]$, $i\in\{\tau,\nu,\theta\}$, where the reference grid intervals are $h_{\tau,0}=130.21$ ns, $h_{\nu,0}=1.875$ kHz, and $h_{\theta,0}=0.6^\circ$.  
The proposed SBL procedure refines at most $K_{\mathrm{act}}=3$ grid points with $\rho=10^{-3}$, peak-exclusion radius $1.75$ DDA bins, and $u_{\max}=0.75$. The Armijo backtracking uses $c_S=10^{-4}$, $\beta=0.5$, and minimum step $2^{-12}$. Damping is set to $10^{-5}\max\{1,\max_i|[\mathbf D_t^T\mathbf H_t\mathbf D_t]_{i,i}|\}$; the stopping tolerances are $\varepsilon_{\mathcal L}=5\times10^{-5}$ and $\varepsilon_\eta=5\times10^{-3}$, with at least five EM iterations. 

The RMSE and NMSE are employed as the performance metrics to quantify the performance of the proposed off-grid SBL framework, which are defined as 
\begin{align}
\mathrm{RMSE}_x&=\sqrt{\frac{1}{N_{\mathrm{MC}}L}
\sum_{m=1}^{N_{\mathrm{MC}}}\sum_{l=1}^{L}
(\hat x_{m,l}-x_{m,l})^2},
\label{eq:rmse}\\
\mathrm{NMSE}_{h}&=\frac{1}{N_{\mathrm{MC}}}\sum_{m=1}^{N_{\mathrm{MC}}}
\frac{\|\hat{\mathbf h}_m-\mathbf h_m\|_2^2}{\|\mathbf h_m\|_2^2},
\label{eq:nmse}
\end{align}
where $x\in\{\tau,\nu,\theta\}$ and $N_{\mathrm{MC}}$ denote the number of Monte-Carlo realizations.

\subsection{RCS and target parameter estimation versus SNR}
Fig.~\ref{fig:rcs} compares the NMSE performance of the RCS coefficient estimates of different schemes. One can observe that the proposed off-grid SBL scheme achieves lower NMSE than on-grid SBL, OMP, and focal underdetermined system solver (FOCUSS). This is because the on-grid schemes approximate the response of an off-grid target using neighboring dictionary columns. Thus, the RCS estimate is affected by both additive noise and mismatch in the selected target coordinates. As the SNR increases, the noise contribution decreases while the mismatch remains. The proposed continuous refinement reduces this mismatch and, consequently, improves the RCS coefficient estimates. The poor performance of OMP is due to the stopping criterion and the choice of the dictionary matrix, while the FOCUSS scheme is sensitive to the choice of the regularization parameter.
Next, Figs.~\ref{RangeRMSE}, \ref{VelocityRMSE} and \ref{AngleRMSE} present range, velocity and angle RMSE versus SNR performance of different schemes, respectively. As expected, the off-grid SBL algorithm outperforms the other benchmark schemes as they fail to account for off-grid targets. The proposed off-grid SBL procedure refines the target coordinates within their bins and achieves decreasing estimation errors. The delay and Doppler RMSEs follow the decreasing root Cramér–Rao bounds (CRBs) more closely at higher SNR, while the angular RMSE retains a larger gap.

\subsection{Effect of DDA Grid Coarsening}
To examine sensitivity to the initial grid, the spacings are increased as $h_i(q)=q h_{i,0}$, where $q\in\{1,1.5,2,3,4\}$ and $i\in\{\tau,\nu,\theta\}$. The search limits and target realizations are fixed across $q$, while the initial dictionary contains $4225$, $1377$, $637$, $225$, and $112$ columns, respectively. The realizations occupy distinct nearest DDA bins and satisfy at least two-bin normalized separation on the coarsest grid.  The joint normalized DDA RMSE is defined as
\begin{align}
\mathbf e_{m,l}&=\left[
\frac{\hat\tau_{m,l}-\tau_{m,l}}{h_{\tau,0}},
\frac{\hat\nu_{m,l}-\nu_{m,l}}{h_{\nu,0}},
\frac{\hat\theta_{m,l}-\theta_{m,l}}{h_{\theta,0}}\right]^T,\nonumber\\
\mathrm{RMSE}_{\mathrm{DDA}}&=
\sqrt{\frac{1}{N_{\mathrm{MC}}L}
\sum_{m=1}^{N_{\mathrm{MC}}}\sum_{l=1}^L\|\mathbf e_{m,l}\|_2^2}.
\label{eq:joint}
\end{align}
Fig.~\ref{fig:grid} demonstrates that the proposed off-grid SBL scheme achieves lower joint RMSE than the on-grid benchmarks at every grid interval. Its accuracy is less sensitive to moderate grid coarsening, since continuous refinement compensates for the increased coordinate quantization. One can observe that, at $q=3$, the initial dictionary contains $225$ columns, compared with $4225$ columns for $q=1$. Despite the increased DDA grid intervals, the proposed off-grid SBL scheme maintains lower joint normalized DDA RMSE than the on-grid benchmark schemes. These results demonstrate that continuous DDA refinement results in improved estimation accuracy using a smaller initial dictionary.
\section{Conclusion}
\label{sec:conclusion}
An off-grid SBL framework was developed for joint DDA and RCS estimation in OTFS-based mmWave MIMO radar. Initially, a sparse 3D parameter estimation model is developed incorporating fractional delay and Doppler shifts. Subsequently, an EM-based procedure estimates the sparse coefficients and noise variance while refining the DDA coordinates through a joint damped GN update. This refinement reduced basis mismatch without requiring a finer DDA grid. Furthermore, Armijo backtracking evaluated each bounded update using the exact continuous dictionary and accepted only updates that decreased the posterior-averaged reconstruction error, thus maintaining the generalized EM property.
\bibliographystyle{IEEEtran}
\bibliography{myRef.bib}
\end{document}